# Chemical assembly of atomically thin transistors and circuits in a large scale

Mervin Zhao[1,2†], Yu Ye[1,2†], Yimo Han[3†], Yang Xia[1†], Hanyu Zhu[1], Yuan Wang[1], David A. Muller[3,4], Xiang Zhang[1,2*]

[1]NSF Nanoscale Science and Engineering Center, University of California, Berkeley, CA 94720, USA.

[2]Materials Sciences Division, Lawrence Berkeley National Laboratory, Berkeley, CA 94720, USA.

[3]School of Applied and Engineering Physics, Cornell University, Ithaca, New York 14853, USA.

[4]Kavli Institute at Cornell for Nanoscale Science, Ithaca, New York 14853, USA.

† These authors contributed equally to this work.

* Correspondence and requests for materials should be addressed to X. Z. (email: xiang@berkeley.edu)

**Abstract:** Next-generation electronics calls for new materials beyond silicon for increased functionality, performance, and scaling in integrated circuits. Carbon nanotubes and semiconductor nanowires are at the forefront of these materials, but have challenges due to the complex fabrication techniques required for large-scale applications[1-4]. Two-dimensional (2D) gapless graphene and semiconducting transition metal dichalcogenides (TMDCs) have emerged as promising electronic materials due to their atomic thickness, chemical stability and scalability. Difficulties in the assembly of 2D electronic structures arise in the precise spatial control over the metallic and semiconducting atomic thin films. Ultimately, this impedes the maturity of integrating atomic elements in modern electronics[5]. Here, we report the large-scale spatially controlled synthesis of the single-layer semiconductor molybdenum disulfide ($MoS_2$) laterally in contact with conductive graphene. Transition electron microscope (TEM) studies reveal that the single-layer $MoS_2$ nucleates at the edge of the graphene, creating a lateral 2D heterostructure. We demonstrate such chemically assembled 2D atomic transistors exhibit high transconductance (10 µS), on-off ratios (~$10^6$), and mobility (~20 cm$^2$ V$^{-1}$ s$^{-1}$). We assemble 2D logic circuits, such as a heterostructure NMOS inverter with a high voltage gain, up to 70, enabled by the precise site selectivity from atomically thin conducting and semiconducting crystals. This scalable chemical assembly of 2D heterostructures may usher in a new era in two-dimensional electronic circuitry and computing.

Moore's law projects that integrated electronics will soon require sub-10 nm transistors. With silicon, this task becomes extremely challenging as the transistor's channel thickness becomes greater than the channel's length, ultimately leading to difficult electrostatic control via the transistor gate[5]. The chemical synthesis of nanomaterials such as inorganic nanowires[2, 3, 6] and carbon nanotubes[4, 7, 8] were aimed at addressing this issue, but utilizing them in electrical devices requires precise placement and orientation using complex fabrication techniques[5, 6, 8, 9]. While great strides have been made, such as the demonstration of a carbon nanotube computer[10], the fabrication to create sophisticated electronic circuitry using these materials remains difficult or impractical. Recently emerging crystalline two-dimensional materials, pose an elegant solution to the problematic scaling in silicon transistors with their ultimate atomic thickness. Graphene, the first widely studied two-dimensional crystal, is a semi-metal with a massless carrier dispersion, high mobility, and easily tunable Fermi level. While single-layer graphene lacks an electronic band-gap, rendering it unsuitable for transistor channels (as it cannot easily be turned "off"), the excellent conductive properties make it ideal for the interconnections and wiring of next-generation devices[11-14].

Recently, semiconducting TMDCs, such as $MoS_2$, has found success as a single-layer transistor[15, 16]. While early works using single-layer TMDCs relied on the "Scotch tape" method of micromechanical exfoliation, considerable efforts have established the chemical vapor deposition (CVD) of various TMDCs[17-21]. For example, CVD grown lateral heterostructures which utilize two different TMDCs have been able to create atomically sharp *p-n* junctions[22, 23]. While lateral heterostructures using wide-gap insulators and conductors, h-BN and graphene, have also been grown[24, 25], there has yet been a demonstration of the spatially controlled synthesis of conductor-semiconductor 2D heterostructures, a necessary step towards full atomically thin circuitry.

Here, we chemically assemble two-dimensional lateral junctions using graphene-$MoS_2$-graphene heterostructures, as shown in Figure 1. Unlike previous reports which rely on transferring and physically assembling transistors using these two 2D crystals[26-28], we chemically grow these transistors in a large scale. By effectively injecting current from the graphene through the $MoS_2$, we demonstrated an NMOS inverter for logic operations, using such 2D heterostructure transistors.

To chemically assemble our heterostructures, single-layer graphene is first transferred onto a silica substrate, as the growth and transfer techniques of graphene are now commonly in scale[11, 12] (Fig. 1a). Patterns of channels (we have chosen a grid) are defined via oxygen plasma (Fig. 1b). The graphene on silica is then placed into a quartz tube for the site selective CVD of single-layer $MoS_2$. The graphene edges facilitate a high density of $MoS_2$ nucleation and the selective growth on the $SiO_2$ channels results in the merging of individual domains which forms a continuous, polycrystalline single-layer of $MoS_2$, similar to the observations made in large area CVD growth on bare substrates (Fig. 1c)[17, 20, 21]. (See methods for more growth details.)

The heterostructure's coverage can be millimeter in scale, as shown in Fig 2a, where the uniform single-layer $MoS_2$ is observed within defined channels from etched graphene. From optical characterizations, single domain $MoS_2$ triangles are observed along the graphene edges, suggesting the preference of nucleation (see Supplementary Information). In addition, brighter regions observed along the junction between graphene and $MoS_2$ (Fig. 2b) may be the nucleation center of single-crystal domains of $MoS_2$, similar to previous reports[19].

Using photoluminescence mapping, the heterostructures displayed strong photoluminescence emission centered at 660 nm between the graphene areas suggesting the presence of a direct band-gap semiconductor (Fig. 2c). Raman mapping (Fig. 2d) further confirmed the grown material is single-layer $MoS_2$ (via the $E_{2g}$ and $A_{1g}$ $MoS_2$ Raman modes[29]), and does not show appreciable intensities on the graphene contacts. Additionally, the graphene Raman signature is preserved after the growth (Fig. 2e), as seen from either the G or 2D peak[30]. (Representative full spectra of the regions can be found in the Supplementary Information).

We evaluate the crystallinity of the grown single layer $MoS_2$ as well as the atomic edge between $MoS_2$ and graphene using transmission electron microscopy (TEM)[18, 19, 31]. The bright-field and dark-field TEM (BF- and DF-TEM) images in Fig. 3a-b are used to create a false-color map of the heterostructure. Mapping red to $MoS_2$ and yellow to graphene, an orange colored line is observed in the overlapped region, indicating a finite overlap of the two crystalline layers. Coupled with optical microscopy images, the $MoS_2$ grain size inside the channel can range from 1 to roughly 10 μm inside the channel (Supplementary Information).

To further investigate the chemistry of the graphene-$MoS_2$ junction, an electron energy loss spectroscopy (EELS) map and integrated EELS spectrum (Fig. 3c) provide the compositional and bonding nature within the heterostructure. From the EELS spectrum, the carbon K-edge is used to identify graphene, while distinct Mo and S edges can be used for $MoS_2$. Within the junction, both signatures of graphene and $MoS_2$ can be observed, confirming a van der Waals heterojunction structure without the formation of additional covalent bonding. The annular dark field STEM (ADF-STEM) images shown in Fig. 3d-e are high-resolution images of the graphene-$MoS_2$ junction. From Fig. 3d, a line profile is created to show a clear overlapped junction due to the atomic number dependence of the ADF-STEM image intensity (Fig. 3f).

Using the ADF-STEM images, nucleation can be confirmed to occur along the graphene edges, as seen in the growth of additional triangular second layer and multilayer $MoS_2$, shown as the bright triangular features in Fig. 3d. Additionally, ADF-STEM is able to resolve the atomic interface between the graphene and $MoS_2$ (Fig. 3e), showing the junction between the graphene and $MoS_2$ does not contain many crystalline defects and may be atomically clean.

The role of graphene as growth mask is surprising, as the nucleation of $MoS_2$ (a three-atom-thick crystal), is able to successfully grow outwards from a single atomic layer. Thus far, lateral heterostructures between van der Waals crystals have shared similar lattice constants. Heterostructures grown using two different TMDCs show that there is typically a preference for lateral epitaxy which is accompanied by the vertical growth as growth time increases[32]. In the case with graphene and $MoS_2$, the lattice mismatch (in- and out-of-plane), as well as surface chemistry difference inhibits the lateral epitaxy. From the TEM studies, $MoS_2$ is observed to nucleate from the edges (Fig. 3d) and forms a junction with a nanometer-scale sized overlap of graphene and $MoS_2$, suggesting a significant reduction in the nucleation energy barrier of $MoS_2$ at the edges of graphene, compared to graphene's surface. After nucleating at the edge, the surface chemistry difference allows $MoS_2$ to preferentially grow on the $SiO_2$, with minimal growth occurring on the graphene surface. This is supported by the fact that the vertical epitaxial growth of $MoS_2$ on graphene is difficult to create large crystalline areas and requires unique conditions not present in this growth[33]. Furthermore, the hydrophilic $SiO_2$ surface (due to the oxygen plasma treatment to etch graphene) has a higher affinity for the growth seed's mass transport, especially compared to the hydrophobic graphene surface[21] contributing to the preferential growth. Critically, graphene acts as a metallic electrode to inject current into the $MoS_2$ in this heterostructure. Transistors fabricated using physically transferred graphene as the electrode to MoS$_2$ have demonstrated that it is an efficient electrical contact material, as graphene's Fermi level can be tuned[26-28]. This has potential to create ohmic contact transistors, an inherent challenge in the use of MoS$_2$ due to the typical Schottky-limited transport[34]. Hence, the nucleation along the graphene edges as well as the chemistry of the junction is expected to play a crucial role in the electrical transport characteristics of these heterostructure transistors.

The field-effect transistor (FET) performance of the graphene-MoS$_2$-graphene structure is measured at room temperature (optical images of the devices are in the supplemental material). The *I-V* curves show linear behaviors at small source-drain voltages, confirming the ohmic contact between the graphene and the single-layer MoS$_2$ film (Fig. 4a). In addition, the source-drain current saturates at a larger source-drain voltage (1.5 V), a crucial parameter for reaching maximum possible operating speeds and maximizing the intrinsic transistor gain[13]. Through electrical transport measurements, the heterostructure exhibits typical *n*-type channel characteristics. Figure 4b shows the top-gate dependence for the heterostructure FET under bias voltages of 1 V. The turn-on voltage is measured at around −1.5 V, suggesting a relatively high electron-doping concentration. These measurements show that the heterostructure exhibits a high on/off current ratio of ~$10^6$, peak transconductance close to 10 μS, and corresponding carrier concentration of $5.6 \times 10^{12}$ cm$^{-2}$. Using the gate dependence, the room temperature field-effect electron mobility is determined to be over 20 cm$^2$ V$^{-1}$ s$^{-1}$, higher than the mobilities from typical exfoliated and chemically-grown MoS$_2$ devices[20]. Hence, we demonstrate that the chemically synthesized MoS$_2$ contacted with graphene electrodes allows for excellent electrical properties and transistor performance.

Furthermore, 2D atomic logic circuitry of an inverter (a NOT gate) (Fig. 5a) is assembled using the lateral heterostructures. Figure 5b shows the static voltage transfer characteristics of the heterostructure NMOS inverter with driving voltages of 1, 2, and 4 V with clear inverting behavior. At low $V_{in}$, starting at −2.5, $V_{out}$ is consistently at the level of $V_{DD}$. Once $V_{in}$ reaches around reaches a threshold of −1.75 V, $V_{out}$ quickly switches to zero and maintains this voltage up to $V_{in} = 0$ V. The heterostructure inverter also yields an extremely high voltage gain, reaching 70 for $V_{DD} = 4$ V (Fig. 5c), one of the highest among all the inverter gates made from TMDC materials[35-37]. As we demonstrate the ability of logic, having control over arbitrary designed patterns may potentially enable the development of 2D circuitry for computing

In conclusion, we have successfully demonstrated chemically assembled atomically-thin lateral heterostructures composed of conductive graphene and semiconducting $MoS_2$. Large-area coverage, over a millimeter in size, and functionality is achieved due to the dual nature of graphene as a growth mask and electrical contact. We show these heterostructures form transistors with on-off ratios of $10^6$ and mobilities greater than 20 $cm^2\,V^{-1}\,s^{-1}$. These transistors are used to form inverters that have voltage gains up 70. It is envisioned that these fully 2D heterostructures may be chemically assembled to form ballistic transistors in the short-channel limit with high performance, enabling us to develop alternatives to silicon technologies. Coupled with the industrial wafer-scale compatibility of graphene and $MoS_2$ growths, the precise spatial control over the synthesis of 2D conductor-semiconductor heterostructures paves the way for next-generation electronics and computing.

**Methods**

**Graphene fabrication:**

Wafer-scale wet-transferred graphene was purchased from Graphenea. Photolithography is used to define a variety of channel widths in the graphene, using PMMA and I-line as a photoresist.

Oxygen plasma is then used to etch away the opened channels in the graphene. The photoresist is washed away using acetone and then annealed in an Ar/H$_2$ environment for two hours at 300 °C.

**Heterostructure growth:**

The heterostructure growth was performed following a procedure similar to reports of seeded CVD MoS$_2$ growth[21]. The chemical seed PTAS (perylene-3,4,9,10-tetracarboxylic acid tetrapotassium salt) was purchased from 2D Semiconductors and dispersed into water to create a solution corresponding to approximately 20 μM, 5 μL of the solution was placed onto the substrate with graphene and blown off using nitrogen. Approximately 18 mg of solid molybdenum trioxide powder was placed in an alumina crucible at the center of the furnace while 16 mg of solid sulfur was placed upstream at a lower temperature. The substrate containing the graphene patterns was placed directly above the MoO$_3$ powder. The temperature was ramped to 650 °C at a rate of 15 °C/min and kept at 650 °C for five minutes. High purity argon was used as a carrier gas to first flush the tube for five minutes, and a rate of 5 SCCM was used during the growth. Following the growth, the furnace was opened and argon was flushed at a high flow rate to rapidly cool the sample. Samples were optically characterized using a Raman microscope set up (Horiba LabRAM HR Evolution) with a 473 nm excitation laser.

**TEM sample preparation:**

The sample was coated with polypropylene carbonate (PPC). The Si/SiO$_2$ substrate was etched by 1M KOH solution at 90°C. Then the film supported by PPC was rinsed in deionized water for three times, 10 min for each, and transferred to a QUANTIFOIL holey carbon TEM grid. Afterwards we baked the sample in vacuum ($10^{-7}$ Torr) at 350°C for 5 hours to remove the PPC. The sample was baked again in an ultrahigh vacuum bake-out system for 8 hours at 130°C to further clean polymer residue before loading into STEM.

**TEM, STEM and EELS:**

BF and DF-TEM images were taken using an FEI Tecnai T12 Spirit operated at 60 keV. ADF-STEM images were taken using a Nion Ultra STEM 100 operated at 60 keV. The beam convergence angle was 35 mrad, and the probe current was ~60 pA. The acquisition time was 32 μs per pixel. The EELS spectrum and map were acquired using a Gatan Quefina dual-EELS spectrometer with an energy dispersion of 0.1 eV/channel. The EELS false-color elemental map was created by integrating the C-K edge and S-$L_{2,3}$ edge. The imaging and EELS mapping conditions are similar to those in other reports[19, 20, 38, 39]. All the EELS analysis was done with open-source Cornell Spectrum Imager software.[40]

**Electrical measurements:**

For the fabrication of field-effect transistors, Pd/Au (10/80 nm) contacts on the graphene for current injection were defined by electron-beam lithography, followed by electron beam evaporation. An atomic layer deposition (ALD) of 20 nm $ZrO_2$ is used for the top-gate dielectric, afterwards an additional top-gate electrode (Pd/Au) is defined on top of the $MoS_2$ channel.

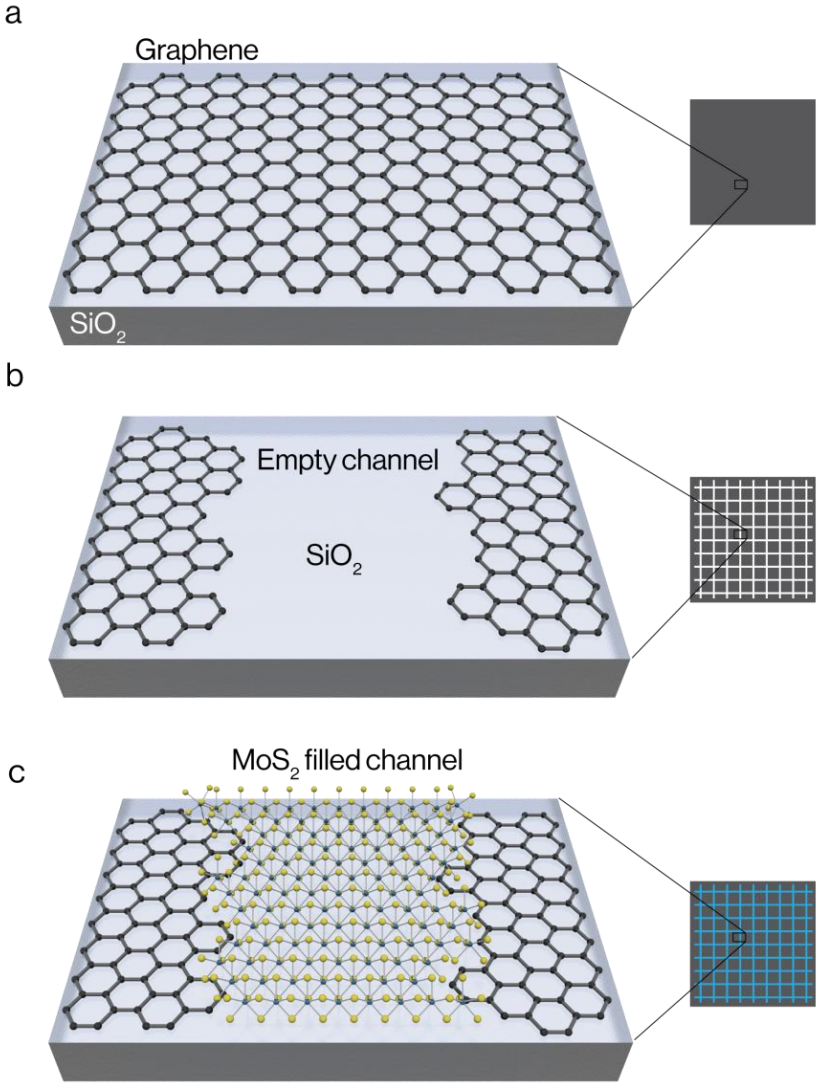

**Figure 1 | Schematic of the graphene-MoS$_2$ heterostructure assembly. a,** To achieve a large-scale growth, graphene is first wet transferred onto a silica substrate. **b,** The graphene is defined into channels using an ultraviolet (UV) photolithography and plasma etching. The entire substrate has patterns defined, as seen in the zoomed-out inset. **c,** Placing the graphene substrate into a furnace allows for growth of polycrystalline MoS$_2$ along the graphene's edges using chemical vapor deposition (inset: zoom-out of MoS$_2$, blue, filled channels between graphene). The growth results in the filling of the graphene channels via a lateral heterostructure that is electrically contacted along graphene's edges.

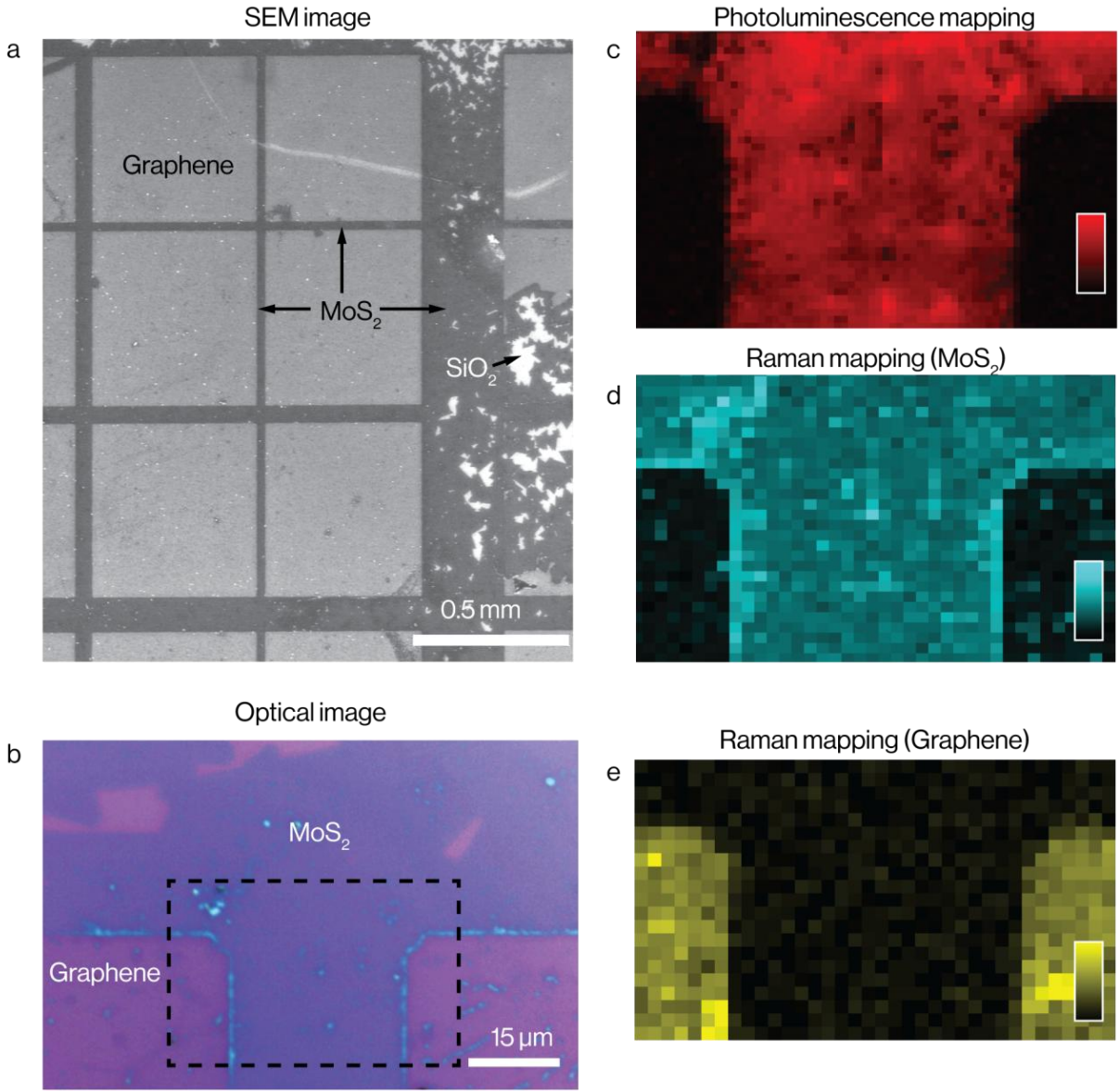

**Figure 2 | Images and spectral characterizations of graphene-MoS$_2$ heterostructure. a,** Scanning electron microscope image of the chemically grown MoS$_2$ between the graphene's edges. The image shows a large scale of coverage, millimeter in scale. In areas without graphene, the growth of triangular domains can be seen on the right of the image. **b,** Optical image of the heterostructure. Within the narrow channel, the MoS$_2$ completely fills the area between the graphene. Thicker areas can be observed around the graphene's edges, indicative of the nucleation of the MoS$_2$ areas. The dashed box is the area used for the spectral mapping in the

following figures. **c**, Photoluminescence mapping of the emission from MoS$_2$ centered at 660 nm. MoS$_2$ transitions from an indirect to direct bandgap semiconductor when it is thinned down to a single layer, showing high emission only in areas without graphene. **d**, Raman mapping using the integrated peak intensities from 380 to 415 cm$^{-1}$. MoS$_2$ has two distinct Raman peaks, E$_{2g}$ and A$_{1g}$ centered near 385 and 410 cm$^{-1}$, respectively. These Raman modes support the finding that the crystalline region inside the grown region is indeed MoS$_2$. **e,** Raman mapping using the integrated peak intensities from 1500 to 1700 cm$^{-1}$. Graphene's Raman signature is the G-peak centered near 1600 cm$^{-1}$. The mapping shows that only the contact areas are graphene. The same results can be observed from integrating the intensities around graphene's 2D peak, centered near 2700 cm$^{-1}$.

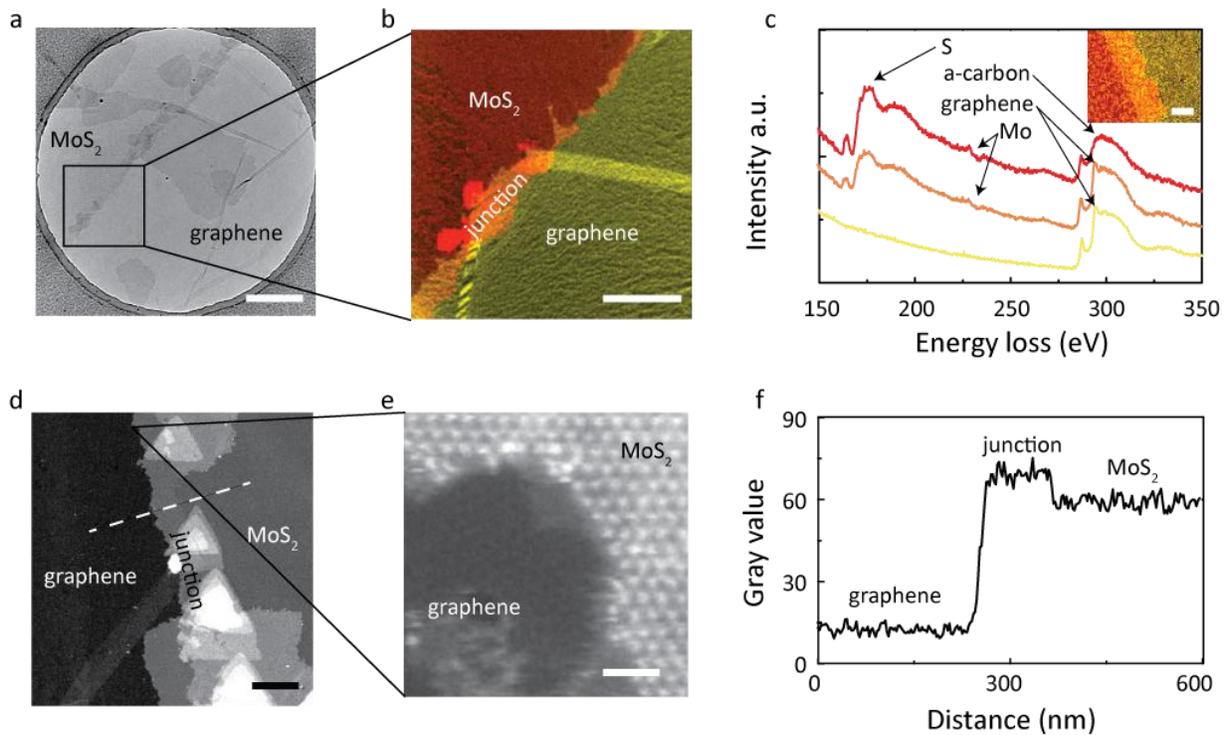

**Figure 3 | Electron microscopy characterizations of the graphene-MoS$_2$ junction. a,** Bright field transmission electron microscope (BF-TEM) images of suspended MoS$_2$-graphene junction. (scale bar: 500 nm) **b,** Dark field TEM (DF-TEM) images compiled to give a false-color map of the MoS$_2$-graphene junction. (scale bar: 200 nm) The graphene region is mapped to a yellow color, while the MoS$_2$ region is mapped to a red color. The overlap junction, shown as an orange color, is a finite overlap between of the two layers. **c,** Electron energy loss spectrum (EELS) of graphene (yellow), MoS$_2$ (red) and the junction (orange) show S, Mo and C edges. The graphene (gr-C) signature in the $\sigma^*$ peak shows up on the graphene and the junction, while only amorphous carbon (a-C) from polymer residue shows up on the MoS$_2$. *Supplemental Figure S5* shows more detailed C-K edges. Inset: EELS map of the junction. (scale bar: 50 nm) Mapping red to MoS$_2$ and yellow to graphene. **d,** Annular dark field scanning transmission electron microscope (ADF-STEM) image of the graphene-MoS$_2$ junction. The white dash line indicates the line profile in (f). (scale bar: 200 nm). The brighter triangles are bilayer or multilayer MoS$_2$, which started to

grow along the graphene edge. **e,** Atomic-resolution ADF-STEM image of graphene-MoS$_2$ junction interface. (scale bar:1 nm) **f**, Image intensity line profile of the graphene-MoS$_2$ junction indicated as dashed line in (d). From the line profile, we are able to distinctly differentiate the graphene, MoS$_2$, and the junction region. From the image intensities the junction is composed of single-layer MoS$_2$ on top of graphene. Detailed 8 bit image gray value calibration is shown in *Supplemental Figure S5*.

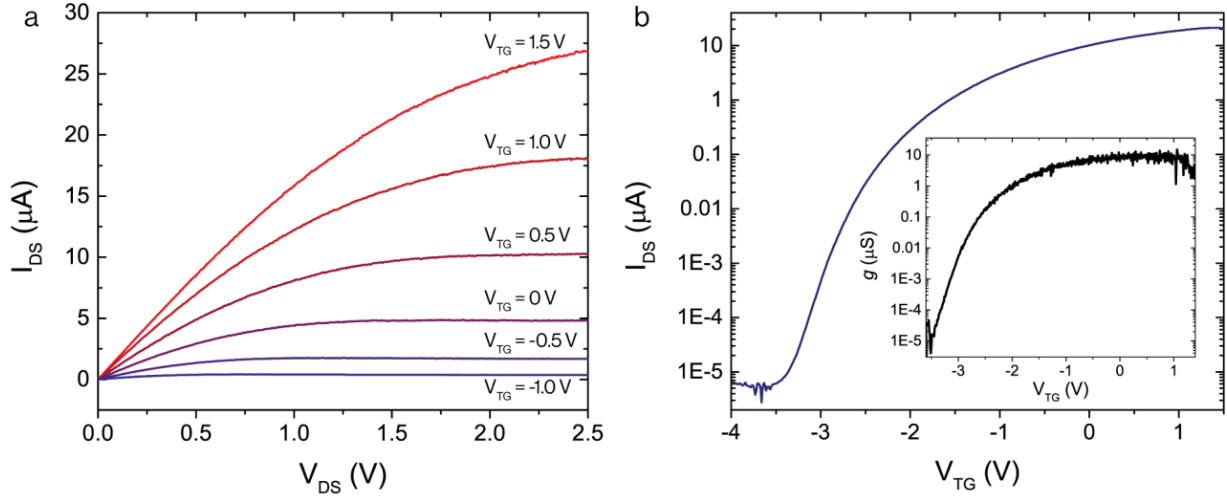

**Figure 4 | Room temperature electrical transport measurements of the graphene-MoS$_2$ heterostructure transistor. a,** Source-drain current curves measured by changing the source-drain bias voltage at differing top-gate voltages. Below a source-drain bias of 1 V, linear ohmic behavior is observed suggesting an efficient electrical contact between the graphene and MoS$_2$. Current saturation is achieved at a bias voltage of 1.5 V. **b,** Source-drain current curves measured by altering the top-gate voltage for a source-drain voltage of 1 V. From the curves an on-off ratio for the heterostructure transistor can be estimated to be approximately $10^6$. The inset shows the transconductance of the heterostructure under $V_{DS}$ = 1 V, reaching a peak of 10 µS. From the curve, the room temperature field-effect mobility is estimated to be ~20 cm$^2$V$^{-1}$s$^{-1}$.

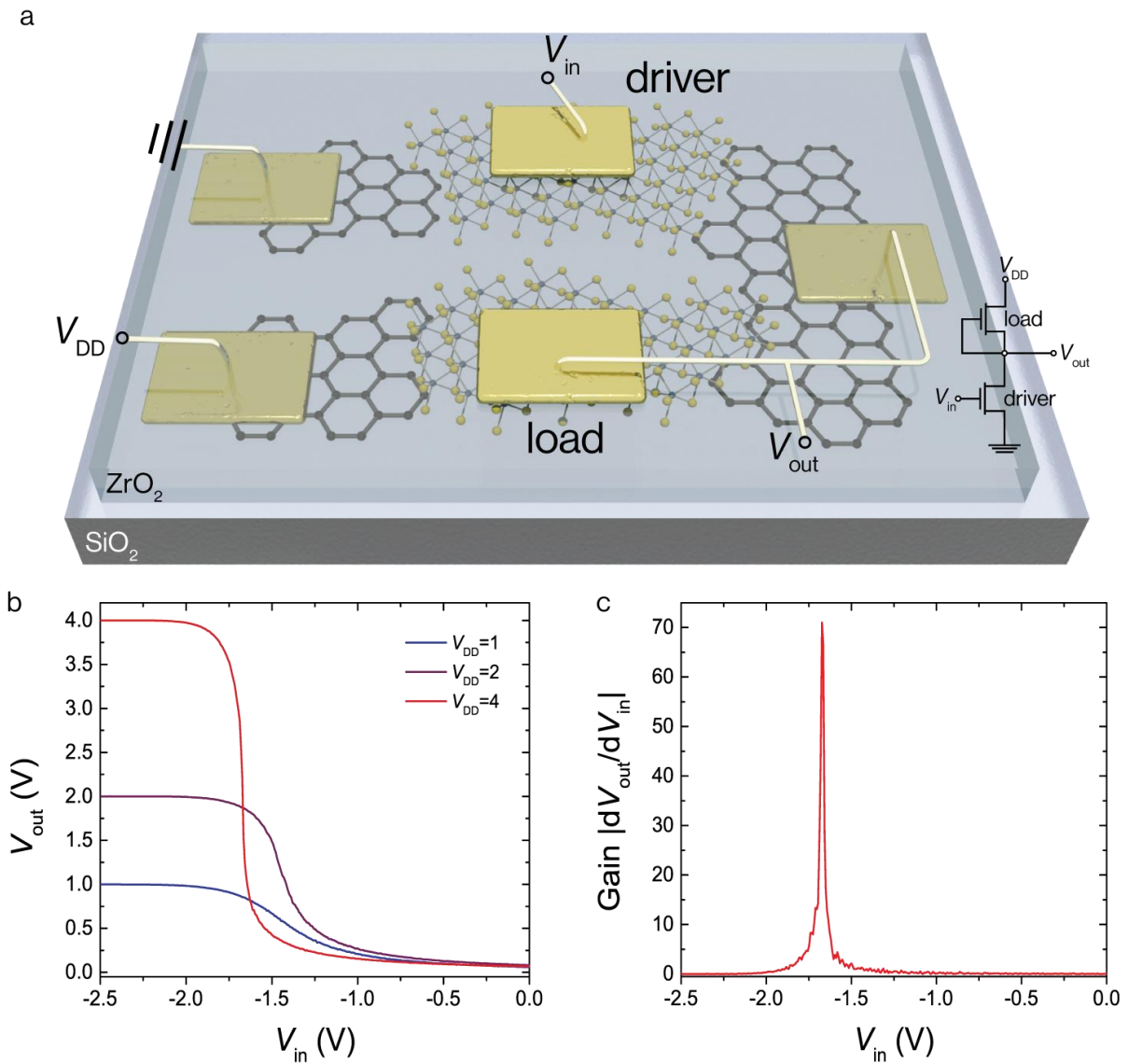

**Figure 5 | Demonstrating logic through a heterostructure inverter. a,** Illustration of the inverter circuit (inset: circuit diagram). Using two *n*-type transistors to create an inverter, the bottom transistor (load) is under a driving voltage while the top transistor (driver) is grounded. The load's gate and the common electrode are read through as the output voltage. The input voltage is applied through the driver's gate. At low input voltages, the driver is turned-off and the output voltage is read as the driving voltage ("1"). At higher input voltages, the system inverts and the output voltage becomes zero, as the driver turns on ("0"). **b,** Inverter transfer

characteristics for driving voltages of 1, 2, and 4 V, showing the inversion behavior at a threshold voltage of around −1.7 V. As described in a., low input voltages yield an output voltage equal to the driving voltage. After crossing the threshold, the output voltage is read as 0. **c,** The absolute differential of the inverter curves gives the voltage gain. At a driving voltage of 4 V, an extremely high voltage gain of 70 is achieved, among the highest in TMDC inverters.

**Supplementary information for "Chemical assembly of atomically thin transistors and circuits in a large scale"**

Mervin Zhao[1,2,†], Yu Ye[1,2,†], Yimo Han[3,†], Yang Xia[1,†], Hanyu Zhu[1], Yuan Wang[1], David A. Muller[3,4], Xiang Zhang[1,2]*

[1]NSF Nanoscale Science and Engineering Center, University of California, Berkeley, CA 94720, USA.

[2]Materials Sciences Division, Lawrence Berkeley National Laboratory, Berkeley, CA 94720, USA.

[3]School of Applied and Engineering Physics, Cornell University, Ithaca, New York 14853, USA.

[4]Kavli Institute at Cornell for Nanoscale Science, Ithaca, New York 14853, USA.

† These authors contributed equally to this work.

* Correspondence and requests for materials should be addressed to X. Z. (email: xiang@berkeley.edu)

**Supplementary Figure S1: Edge-growth**

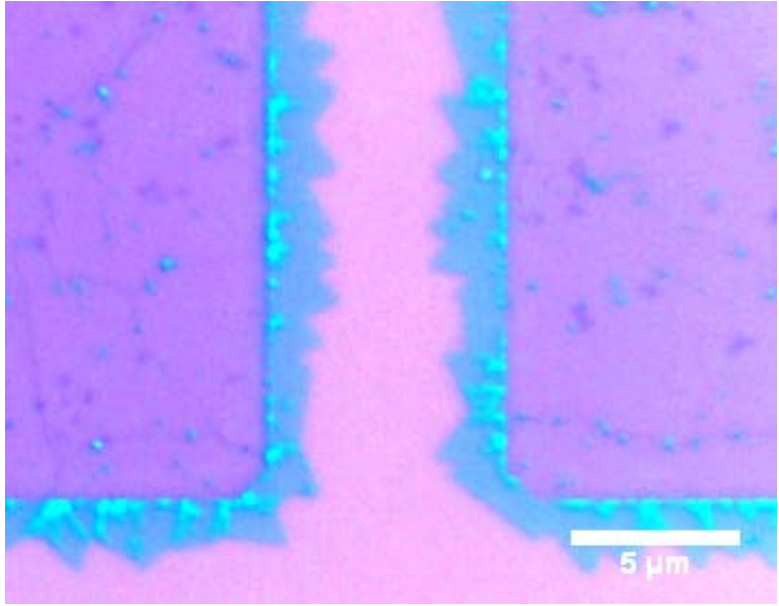

Optical image of the single-layer $MoS_2$ grown from the edge of graphene in an area farther away from the optimized growth area. We can see the nucleation occurs primarily on the edge of graphene and typical triangular shaped $MoS_2$ single layers extend outwards towards the $SiO_2$ surface. Thicker regions corresponding to two or more layers also are localized to the edge, which gives reason to believe that the growth is initiated at the edge.

**Supplementary Figure S2: Growth along irregular graphene areas**

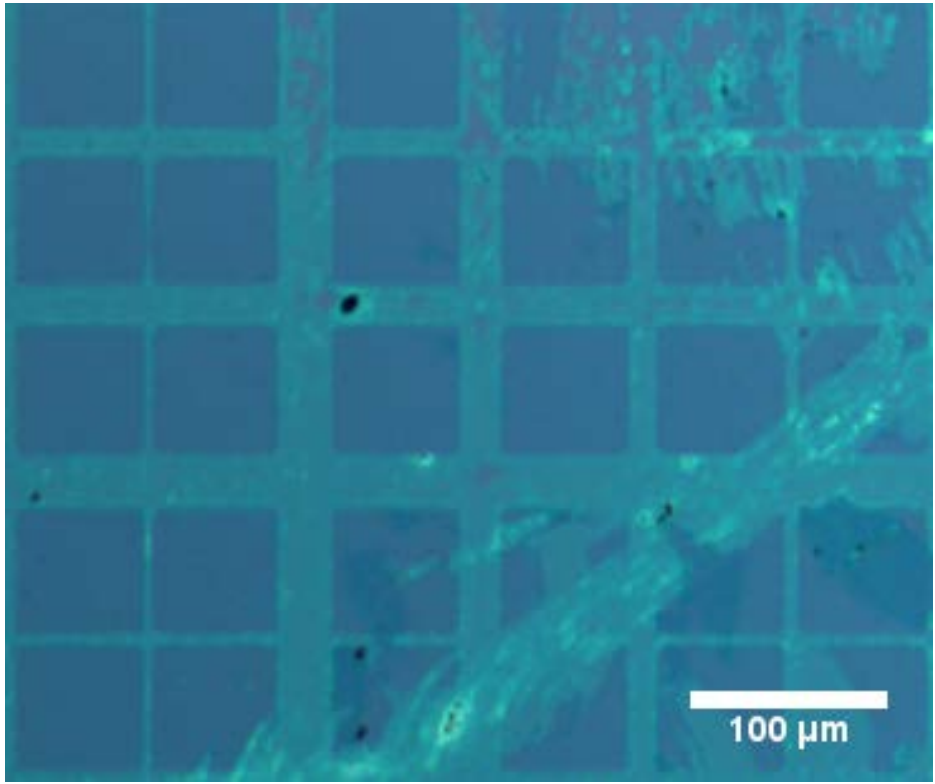

Handling the substrate with graphene with tweezers sometimes will result in areas with damaged graphene as seen in the center and areas on the right. It is seen that single-layer $MoS_2$ is still able to grow from these areas, which have not been through fabrication. This is evidence of the $MoS_2$ nucleation and growth away from the graphene even with irregular edges.

**Supplementary Figure S3: Full spectral information**

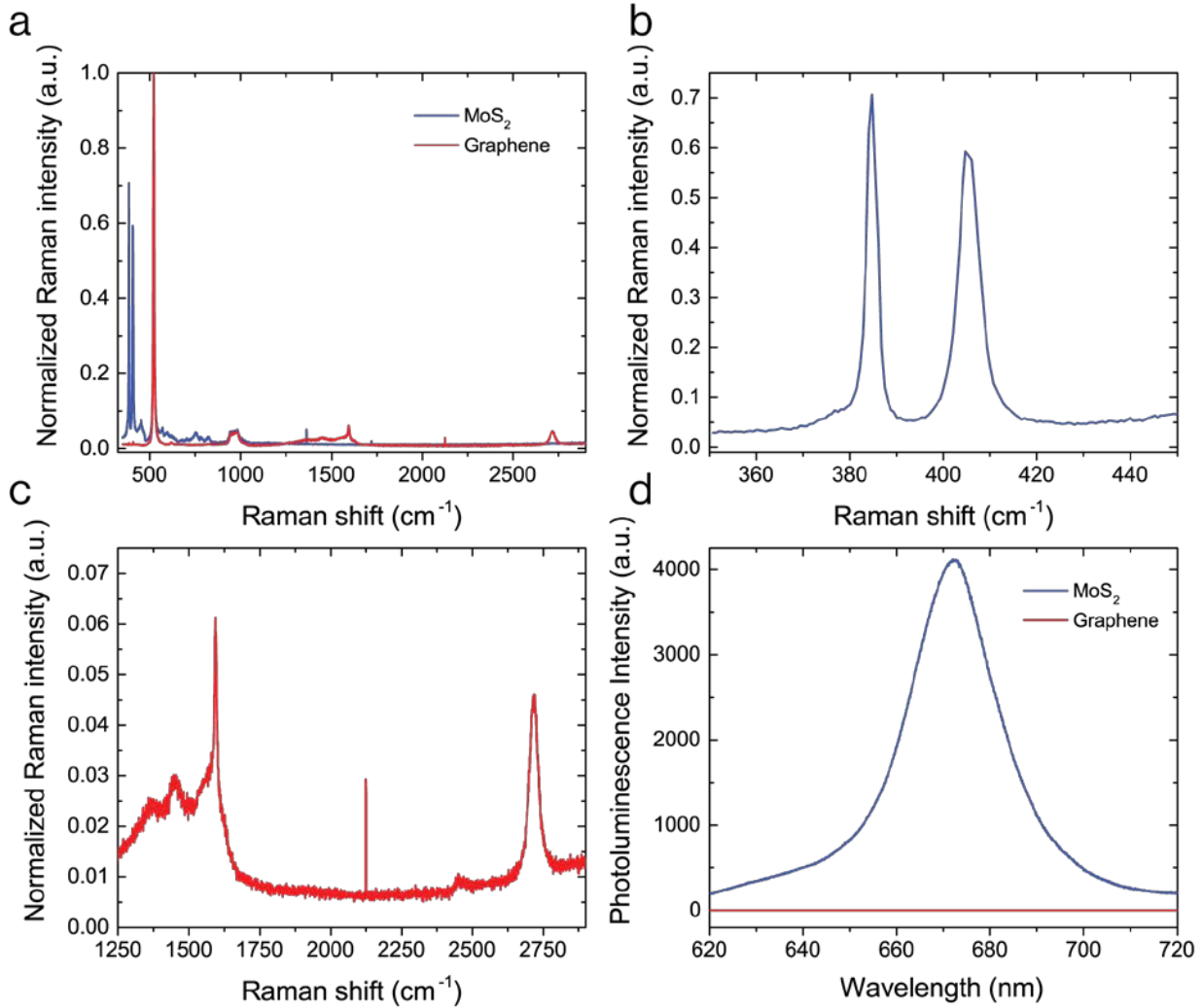

**a,** Full spectra of the MoS$_2$ in the channel and graphene electrodes. We can clearly observe the lack of the MoS$_2$ Raman peaks (**b**) which are lacking in the full spectra as well as vice versa for (**c**). **b,** Zoomed-in spectrum of the MoS$_2$ only which shows the two distinctive Raman peaks given by the literature[1]. The difference in the two peaks is ~20 cm$^{-1}$, which corresponds to single-layer MoS$_2$. **c,** Zoomed-in spectrum of the graphene peaks given by the literature[2]. The Shoulder which appears before the G-peak (1600 cm$^{-1}$) is from the fabrication of graphene followed by annealing. **d**, Photoluminescence (PL) of the MoS$_2$ and graphene areas. Single-layer

MoS$_2$ shows strong emission centered at around 670 nm[3], while the graphene area shows no photoluminescence, further indicative that there is no MoS$_2$ on the graphene areas. In Fig. 3c-e, the mapping is from integration bounds inside a spectral windows. The window is able to cover the entire PL (Fig. 3c). The window for the MoS$_2$ Raman peaks is fully integrated to give Fig. 3d. For graphene, windows are selectively chosen to be centered at the G or 2D peak, and the whole peak is integrated to give Fig. 3e (the G peak).

**Supplementary Figure S4: Transistor images and device performances**

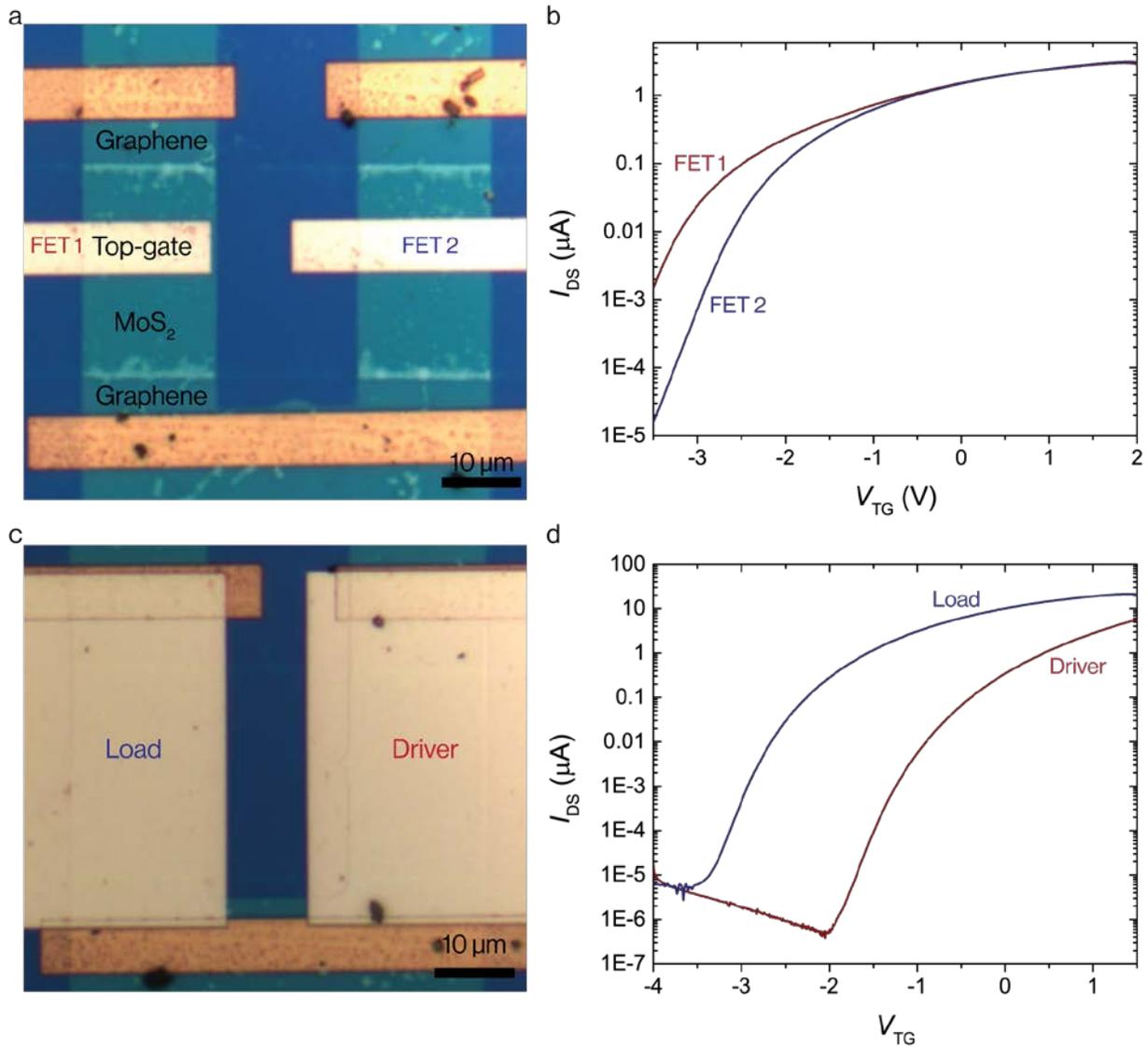

**a,** The heterostructure FETs are already fabricated to resemble inverters (i.e. they share a common electrode), before the top-gate is wire-bonded to the common electrode we have tested the individual performances of two top-gate designations. The source-drain contacts are only defined on the graphene, thus confirming injection of current through graphene into the MoS$_2$. **b,** Using a smaller top-gate on top of only the MoS$_2$ channel areas, we achieve two transfer curves corresponding to the FETs in (**a**) with a more negative turn-on voltage (less than –3 V). **c,** Using a larger top-gate we can slightly contact the graphene as well. The "load" and "driver" here are determined from (**d**) and are used for the load and drivers of the inverter device presented in Figure 5 of the main text. **d,** The larger top-gate configuration has a higher turn-on voltage and the two FETs exhibited higher peak current, making them better suited for the NMOS inverter. However, the differences are not large which indicates in both gate configurations the $MoS_2$ is primarily being tuned to turn-on. The blue transfer curve is the same as that shown in Fig. 4b of the main text.

**Supplementary Figure S5: Electron microscopy and spectroscopy**

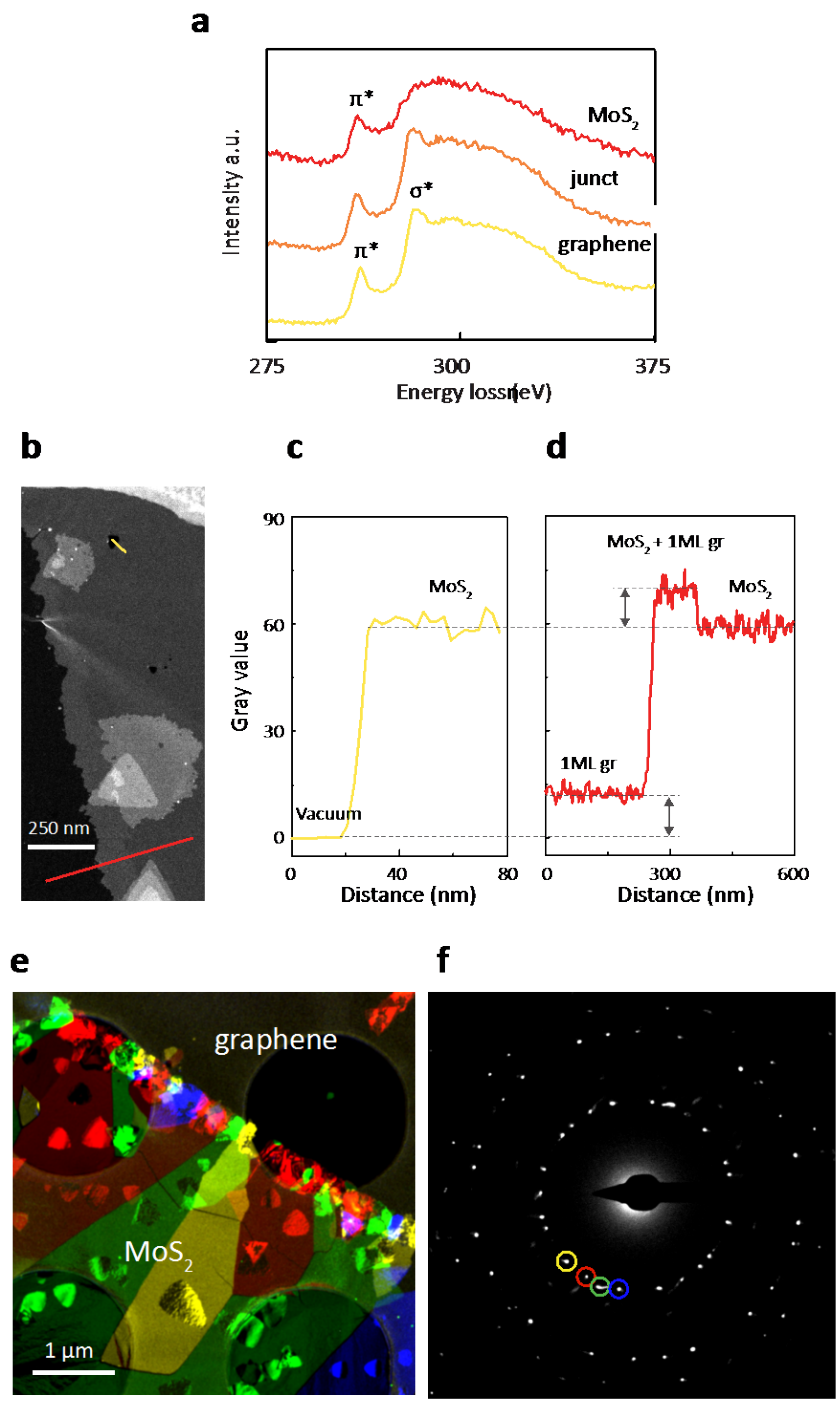

a. Carbon K-edge core-loss spectra from $MoS_2$ (red), the junction (orange) and graphene (yellow). The π* peak appears in both a-C and graphene. The σ* peak in carbon K-edge indicates graphitic carbon. The carbon K-edge from $MoS_2$ contains only a-C signal, which comes from polymer residue. b. HAADF-STEM image with image intensity proportional to $Z^{\gamma}$, where Z is the atomic number and $1.3 < \gamma < 2$. The line profiles of $MoS_2$ edge (c) a $MoS_2$-graphene junction (d) confirmed that the $MoS_2$-graphene junction is an overlapped junction. The DF-TEM (e) and selected-area diffraction pattern (SADP) (f) show the grain sizes and grain orientations of the $MoS_2$ film near the junction.